\documentclass[proc,oneside]{edpsmath} 
\usepackage{graphicx}
\usepackage{url}
\usepackage{tikz}

\usepackage{amsfonts}
\usepackage{mathrsfs}

\usepackage[T1]{fontenc}

\usepackage{multirow}
\usepackage{multicol}
\usepackage[linesnumbered,ruled,vlined]{algorithm2e}
\usepackage{listings}
\usepackage{subcaption}

\usepackage[draft,textwidth=2.5cm]{todonotes} 
\usepackage{soul}
\presetkeys{todonotes}{fancyline}{}

\newcommand{\gysela}{\textsc{gysela}}

\def\be{\begin{equation}}
\def\ee{\end{equation}}

\def\tit{\textit}

\newcommand{\gf}[0]{{\scriptsize GFLOPS}}
\newcommand{\gbs}[0]{{\scriptsize GB/s}}

\newcommand{\gy}[0]{\textsc{Gysela}\xspace}

\usepackage{xspace}
\newcommand{\comment}[1]{}

\newcommand{\intel}{\textsc{Intel}\xspace}

\newcommand{\stream}{\textsc{stream}\xspace}

\newcommand{\vpar}{v_\parallel}

\newcommand{\Bstar}{B_{\|s}^{\ast}}
\newcommand{\bstar}{\vec{b}_s^{\ast}}
\newcommand{\vecbstar}{\vec{b}_s^{\ast}}

\newcommand{\vecb}{\vec{b}}

\newcommand{\rmin}{r_{\rm min}}
\newcommand{\rmax}{r_{\rm max}}

\newcommand{\fbar}{\bar{f}}

\def\be{\begin{equation}}
\def\ee{\end{equation}}

\newcommand{\dv}{\,{\text{d}}{\mathbf{v}}}

\begin{document}

\title{Evaluating kernels on Xeon Phi to\\ accelerate Gysela application}
\author{G.\,Latu}\address{CEA, IRFM, F-13108 Saint-Paul-lez-Durance}%
\author{M.\,Haefele}\address{IPP, Boltzmannstrasse 2, D-85748 Garching }\sameaddress{,3}%
\author{J.\,Bigot}\address{Maison de la Simulation, CEA/CNRS/Inria/Univ. Paris-Sud/Univ. de Versailles, F-91191, Gif-sur-Yvette Cedex}%
\author{V.\,Grandgirard}\sameaddress{1}%
\author{T.\,Cartier-Michaud}\sameaddress{1}%
\author{F.\,Rozar}\sameaddress{1,3}%

\begin{abstract} 
This work describes the challenges presented by porting parts of
the \gysela{} code to the \intel{} Xeon Phi coprocessor, as well as
techniques used for optimization, vectorization and tuning that can be
applied to other applications. We evaluate the performance of some
generic micro-benchmark on Phi versus \intel{} Sandy Bridge. Several
interpolation kernels useful for the \gysela{} application are
analyzed and the performance are shown. Some memory-bound and
compute-bound kernels are accelerated by a factor 2 on the Phi
device compared to Sandy architecture. Nevertheless, it is hard, if
not impossible, to reach a large fraction of the peak performance on the Phi device,
especially for real-life applications as \gysela{}. A collateral
benefit of this optimization and tuning work is that the execution
time of \gy{} (using 4D advections) has decreased on a standard
architecture such as \intel{} Sandy Bridge.
\end{abstract} 

%


\maketitle

\vspace*{-.3cm}
\section{\label{intro}Introduction}
A key factor that determines the performance of magnetic plasma
containment devices as potential fusion reactors is the transport of
heat, particles, and momentum. To study turbulent transport and to model
Tokamak fusion plasmas, several parallel codes have been designed over
the years. In the last decade, the simulation of turbulent fusion
plasmas in Tokamak devices has improved a lot, notably thanks to the
availability of large computers. Computational resources available
nowadays has allowed the development of several petascale codes based
on the well-established gyrokinetic framework.

In this article, we focus on the \gysela{} gyrokinetic code. The relative
efficiency of this application (weak scaling starting from 8k cores) is higher than 95\% at
64k cores on several supercomputers. Producing physics results with
this tool necessitates large CPU resources. Moreover, it is expected
that the needs will increase in the near future (due to high
resolution and addition of kinetic electrons mainly). Adapting the code to new
parallel architectures is then a key issue. It is important to
understand new hardwares, their advantages and limitations.  
This Xeon Phi coprocessor is the first one, largely commercialized, that implements the Many Integrated Cores (MIC) architecture. 
This architecture is appealing, as the compiling and execution 
steps are quite similar to those of standard \intel{} processors, and the theoretical peak 
performance and memory bandwidth are high. Furthermore, we had the opportunity to get 
an access to the Helios machine dedicated
to Fusion community (Japan), where such devices were available. This work
describes the challenges presented by porting parts of the \gysela{}
code to the \intel{} Xeon Phi coprocessor, as well as opportunities for
optimization, vectorization and tuning that can be applied to other
applications.  

Section \ref{mpi_phi} is devoted to the evaluation of
Xeon Phi performance using generic micro-benchmarks. This overview of
the observed performance and the encountered difficulties on Xeon Phi will help us to understand the bottlenecks and benefits of this architecture. Using these information, we will able in following sections to port the \gysela{} application on this device.

To simplify our study, we have extracted
a mini-application from the \gysela{} code with a reduced code line
count, and with less physics submodels. The mini-application is much
easier to modify in order to optimize and to tune the code. 
We tried a \textit{top-down} approach to improve the
performance of the costly computational part located in a single subroutine. Many standard optimization techniques were investigated on this subroutine, but we failed to reach the level of performance that one can expect from Xeon Phi. The description of the mini-application and this first port is given in section \ref{framework}.

Then, we investigated a \textit{bottom-up} approach to overcome the performance limitation we observed. To finely study the optimization aspects, we 
designed very small computation kernels, easy to modify, that are representative of the costly part of the mini-application. The description of these kernels, the optimizations we have done on them and benchmarks results we obtained, are described in section \ref{kernels}. 
Finally, the lesson learned from the porting  of the tiny computation kernels has been used to improve the performance of the mini-application. Section \ref{miniapp_port} describes the end of the \textbf{bottom-up} approach when the optimized kernels are integrated back into the mini-application and the kind of global performance improvement we could observe.


\section{\label{mpi_phi}Generic micro-benchmarks on Xeon Phi}
\subsection{Xeon Phi architecture}
\label{sec:phi_archi}

The benchmarks presented in this document have been mostly obtained on
the Helios machine on IFERC computing facility (Rokkasho, Japan). 
The hardware we used consists of
Xeon Phi cards with 60 cores (1.052 GHz), where each core is capable of
executing four concurrent threads. For comparison, the host processor
(Sandy Bridge, \intel{} Xeon E5-2600) consists of two 2.7 GHz processors 
with eight cores each and with two admissible threads per core.

The types of problems the Xeon Phi is well suited for are intensive
numerical computations.  Additionally, for better performance the
computations can use one of the highly optimized vector math
libraries that were implemented using assembly language constructs
tuned specifically for the Xeon Phi architecture, or very well
vectorized code.

Also, local CPU caches should be used as much as possible.
Maintaining locality of data in caches is a key factor to achieve
performance. This is a major difficulty because the L2 cache is about
25 MB over all 60 cores on the Xeon Phi (and over the possibly 240
threads), which means much less cache memory per core than on the Sandy
Bridge host.

For an exhaustive and detailed architecture review of the Xeon Phi,
the reader can refer
to~\cite{rahman}. Table~\ref{table:chip_comparison} shows the main
hardware features of the Xeon Phi compared to the Sandy Bridge
processor.

\begin{table}[h!]
  \begin{tabular}{|l|c|c|}
    \hline
 \multirow{2}{*}{Processor}  & \intel{} Sandy Bridge (Xeon E5) & \multirow{2}{*}{\intel{} Xeon Phi 5110P}  \\
            & Single socket & \\
    \hline
  Number of cores & 8 & 60 \\
  Available memory & 32 GB & 8 GB \\
  Peak performance (double precision) & 173 GFlops/s & 1 TFlops/s \\
  Sustainable memory bandwidth & 40 GB/s & 160 GB/s \\
  Instruction execution model & Out of order & In order \\
  Clock frequency & 2.7 GHz & 1.05 GHz \\
    L2 Cache size/core & 256 KB & 512KB \\ 
    L3 Cache size & 20 MB & 0 \\ 
    \hline
  \end{tabular}
\medskip
  \caption{Hardware key features for \intel{} Sandy Bridge and \intel{} Xeon Phi chips available on the Helios machine.}
  \label{table:chip_comparison}
\end{table}
One can already notice the impressive factor four in memory bandwidth and factor $5.8$ in peak performance which separates the Sandy Bridge (single socket) from the Xeon Phi.
On the other hand, the memory per core shrinks by a factor 30 from 4GB/core to 130MB/core.
Additionally, the peak performance per core also shrinks from $21.6$ GFlops/s to $16.6$ GFlop/s.
These peak performance numbers both assume the usage of the vectorized fused multiply-add assembly instruction, i.e. one addition and one multiplication are performed on all the elements of the vector registers given in parameter.
We will see in section~\ref{sec:mic_core} that a core of the Xeon Phi should host at least two threads to get a significant fraction of the peak performance, whereas a single thread per core is enough on Sandy Bridge.
As a result, in order to reach a given performance, the parallelization effort is much more important on the Xeon Phi compared to Sandy Bridge.

\subsection{Xeon Phi cluster}
\label{sec:phi_cluster}

\begin{table}
  \begin{tabular}{|c|c|c|c|c|}
    \hline
     Name & Location & \#Nodes & Host & Coprocessor \\
    \hline
    \multirow{2}{*}{Helios MIC} & \multirow{2}{*}{IFERC-CSC, Rokkasho, Japan} & \multirow{2}{*}{180} & 2x Sandy Bridge EP & 2 x Xeon Phi 5110P \\
    & & & 8 cores @2.1GHz & 60 cores @1.05GHz \\
    \hline
    \multirow{2}{*}{Supermic} & \multirow{2}{*}{LRZ, Garching, Germany} & \multirow{2}{*}{32} & 2 x Ivy-Bridge E5-2650 & 2 x Xeon Phi 5110P \\
    & & & 8 cores @2.6GHz & 60 cores @1.1GHz \\
    \hline
    \multirow{2}{*}{Eurora} & \multirow{2}{*}{Cineca, Italy} & \multirow{2}{*}{32} &  2x Sandy Bridge E5-2658 & 2 x Xeon Phi 5110P \\
    & & & 8 cores @2.1GHz & 60 cores @1.1GHz \\
    \hline
    \multirow{2}{*}{Robin} & \multirow{2}{*}{Bull R\&D, Grenoble, France} & \multirow{2}{*}{4} &  2x Sandy Bridge E-2680 v2 & 2 x Xeon Phi 3115A \\
    & & & 8 cores @2.8GHz & 57 cores @1.1GHz \\
    \hline
    \multirow{2}{*}{Mick} & \multirow{2}{*}{RZG, Garching, Germany} & \multirow{2}{*}{1} &  2x Sandy Bridge E5-2680 & 2 x Xeon Phi 7120P \\
    & & & 8 cores @2.7GHz & 61 cores @1.238GHz \\
    \hline
  \end{tabular}
  \caption{Xeon Phi clusters used in the context of the current study}
  \label{table:phi_cluster}
\end{table}

\begin{figure}[htb]
  \includegraphics[height=10cm]{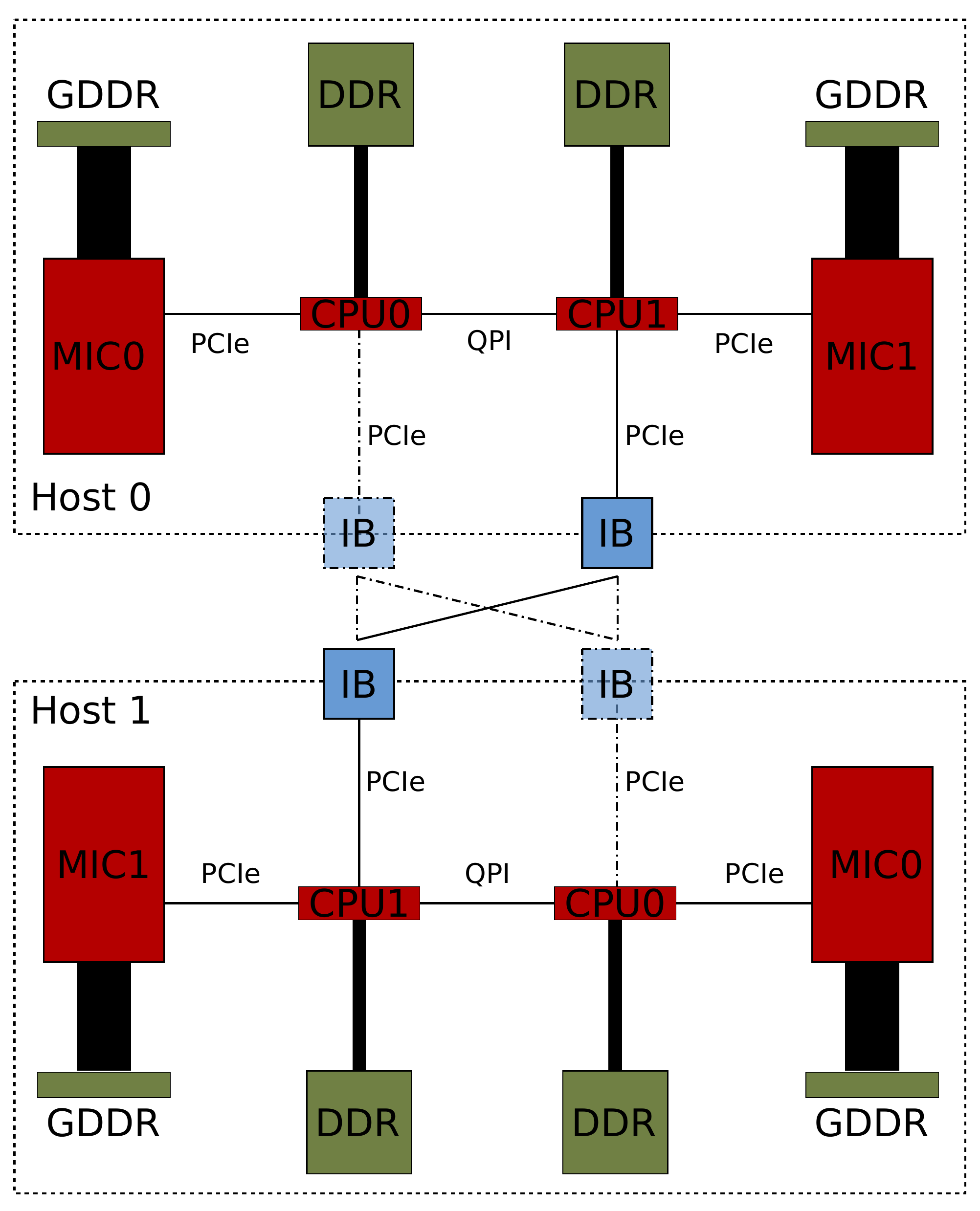}  
  \caption{Architecture of two nodes of a MIC cluster like at IFERC-CSC (HELIOS). Thickness of the lines is proportional to the bandwidth. The size of the boxes is proportional to the size for memory and to the peak performance for computing elements. Dotted lines represents network links available when two network ports are present on each node.}
  \label{fig:phi_node_arch}
\end{figure}

Table~\ref{table:phi_cluster} presents the Xeon Phi clusters used in the context of this study and Figure~\ref{fig:phi_node_arch} depicts two nodes of such a Xeon Phi cluster, having two CPUs and two MICs each.
As one can see on the figure, the Xeon is not connected the same way the main CPUs are.
The Phi embeds its own memory and is connected to the main CPUs using the PCIe bus.
Therefore, the Xeon Phi is regarded as coprocessor or accelerator similarly to GPUs and the main CPU is referred as its host.
Additionally, depending on the hardware, one node can have either one or two InfiniBand (IB in the following) ports. 
Helios MIC and Supermic have two IB ports whereas Robin has only one. 
This information is missing for the Eurora cluster and is not meaningful for Mick as it is a single node.

The current coprocessor design of the Xeon Phi is clearly a limitation and the consequences will be discussed later in section~\ref{sub:network_micro_benchmark}.
Hopefully, this limitation will disappear for the next processor of the MIC family as Knights Landing is expected to have a regular socket.

\subsection{Programming models}
\label{sec:phi_cluster}

The developer has currently two different programming models to implement an application on such architecture.
In the following, these two programming models will be compared regarding three criteria: impact on the code of an existing MPI application, concurrent usage of host and coprocessor resources and availability in production.

The first strategy is the so-called "offload mode", similar to GPUs programming models.
It consists in introducing pragma statements in the code around the computing kernels that should be executed on the coprocessor.
At execution time, when a process running on the CPUs reaches this portion of the code, the execution on the host stops, data are transfered on the coprocessor and the execution resumes on the coprocessor.
Once the execution reaches the end of the pragma section, the execution stops on the coprocessor, data are transfered back on the CPU and the execution resumes on the CPU.
Asynchronous pragmas are available and enable concurrent execution on both the CPU and the coprocessor.
MPI is then used to exploit the resources of several nodes.
The MPI processes running on the CPUs (typically one per socket) implement these pragmas and offload data and computations on the coprocessors.

On the one hand, the advantage of this approach is that it is ready for production today on all the Xeon Phi clusters.
On the other hand, to implement such offload scheme efficiently, the developer has to minimize the data transfer between the host and the coprocessor as it can very easily become a strong bottleneck for the performance.
Additionally, to distribute computations on both the host and the coprocessor at the same time, the usage of asynchronous offload pragmas is required.
This makes the data synchronization between both computation units particularly complex.
Finally, the next processor in the MIC family has been announced by \intel{} to come in a regular ``socket'' form-factor that can be directly used as a main CPU.
This means that the view of the Xeon Phi as a coprocessor might disappear and the offload programming model with it.

The second strategy is the so-called "native" or "symmetric mode", similar to CPUs programming models.
The reason this model is available for Xeon Phi while there is no such model for GPU accelerators is that the Xeon Phi runs a complete operating system whereas GPUs are operated by the system running on the host.
In this model, each coprocessor is seen as a separate node and the MPI application is deployed on both the CPU and MIC nodes.
The largest advantage of this strategy resides in its ease of use: the application developer stays within the well known MPI programming framework.
MPI tasks are independent from each other, run concurrently on different nodes and communicate between them by sending and receiving messages.
However, the set of nodes on which an application runs is no longer homogeneous: there are "host" nodes (Sandy Bridge) and "coprocessor" nodes (Xeon Phi).
The network connecting these nodes is also inhomogeneous (details given in section~\ref{sub:network_micro_benchmark}).
One specific difficulty to this programming model resides in its availability for production. 
Indeed, complex deployment of MPI applications on such heterogeneous hardware requires features that are not yet available in current job schedulers.

For both programming models, the hardware non-homogeneity also leads to possible load imbalance issues.
The Xeon Phi is potentially up to four times more powerful as a single Sandy Bridge with four times less memory.
At the same time a single Sandy Bridge core is much more powerful than a single Xeon Phi core, so serial parts of the code will execute faster on a Sandy Bridge. 
So depending on the computation and for a given load, the execution time will likely be different on the host and on the coprocessor.
This translates differently in the two programming models.
For the native mode, as the large majority of MPI applications distributes equally the same computations with the same load on the different nodes, some tasks will finish earlier than others then leaving their computing unit idle. 
A strategy to alleviate this would consist in modifying the MPI application such that different groups of MPI tasks can have different loads.
But this kind of modifications can be very difficult to implement and is highly application dependent.
For the offload mode, it is much more natural to execute different computations on the host and on the coprocessor.
But a naive implementation uses alternatively the resources of the CPU and the coprocessor letting completely  idle one of the device at any time.
A strategy to alleviate this consists in using the asynchronous pragmas in order to have a concurrent usage of both the host and the coprocessor. 
Of course implementation complexity increases and the load imbalance issue encountered in the native mode has to be solved as well.

As a conclusion on programming models, porting an existing MPI application to get high performance on that kind of architecture is a complex and very time demanding task.
Furthermore, lot's of performance fine tuning on a given architecture will have to be done again on the next hardware generation.
Runtime system like StarPU~\cite{starpu} or XKaapi~\cite{xkaapi} could reduce these costs in the future. 
Indeed, once the developer has expressed the parallelism of his application as a set of tasks, the runtime system schedules them dynamically on the available computing units and performs the required data transfers.
Even if these technologies are becoming more and more mature, there are not yet production ready and still active areas of research.
But in the context of the present study, we focus on the native mode as we plan to use MPI to deploy the final application on a Xeon / Xeon Phi cluster.

\subsection{Vector processing cores micro-benchmarks}
\label{sec:mic_core}
The Sandy Bridge architecture supports the Advanced Vector eXtension (AVX) instruction set that extends the standard x86 instruction set in order to enable vector operations on 256 bits wide vector registers.
This means that a single register can hold eight floating point numbers in single precision or four in double precision.
The Xeon Phi has yet another instruction set that operates on 512 bits wide vector registers (16 single precision or 8 double precision numbers). 
The latency of vector instructions on Xeon Phi is between two and six cycles which is comparable to the Sandy Bridge one's.
However, the instruction throughput is only one instruction every two cycles for a single thread on Xeon Phi where a single thread can issue one instruction every cycle on Sandy Bridge.
As a result, to achieve similar  instruction throughputs one has to run at least
two threads on each core of a Xeon Phi. 
On top of this, the Out-of-order execution feature of the Sandy Bridge allows the execution of the incoming stream of instructions in any order, provided data dependencies are satisfied.
This allows the filling of the different pipelines with instructions for which data are available in cache even if they were issued later.
The simpler in-order execution model of the Xeon Phi leads to empty stages in the pipelines instead.
In order to circumvent this issue, the hardware is ready to welcome up to four threads on each core.
So one has to use additional threads, e.g. three or four threads per core on the Xeon Phi.
This results in a total number of 180 or 240 threads respectively (for a Xeon Phi with 60 cores).
Each core executes alternatively these threads in a round-robin manner in order to hide the instructions, pipelines and memory latencies.

\subsection{Memory micro-benchmarks}
\label{sec:mic_membench}
\begin{figure}
 
  \includegraphics[height=8cm]{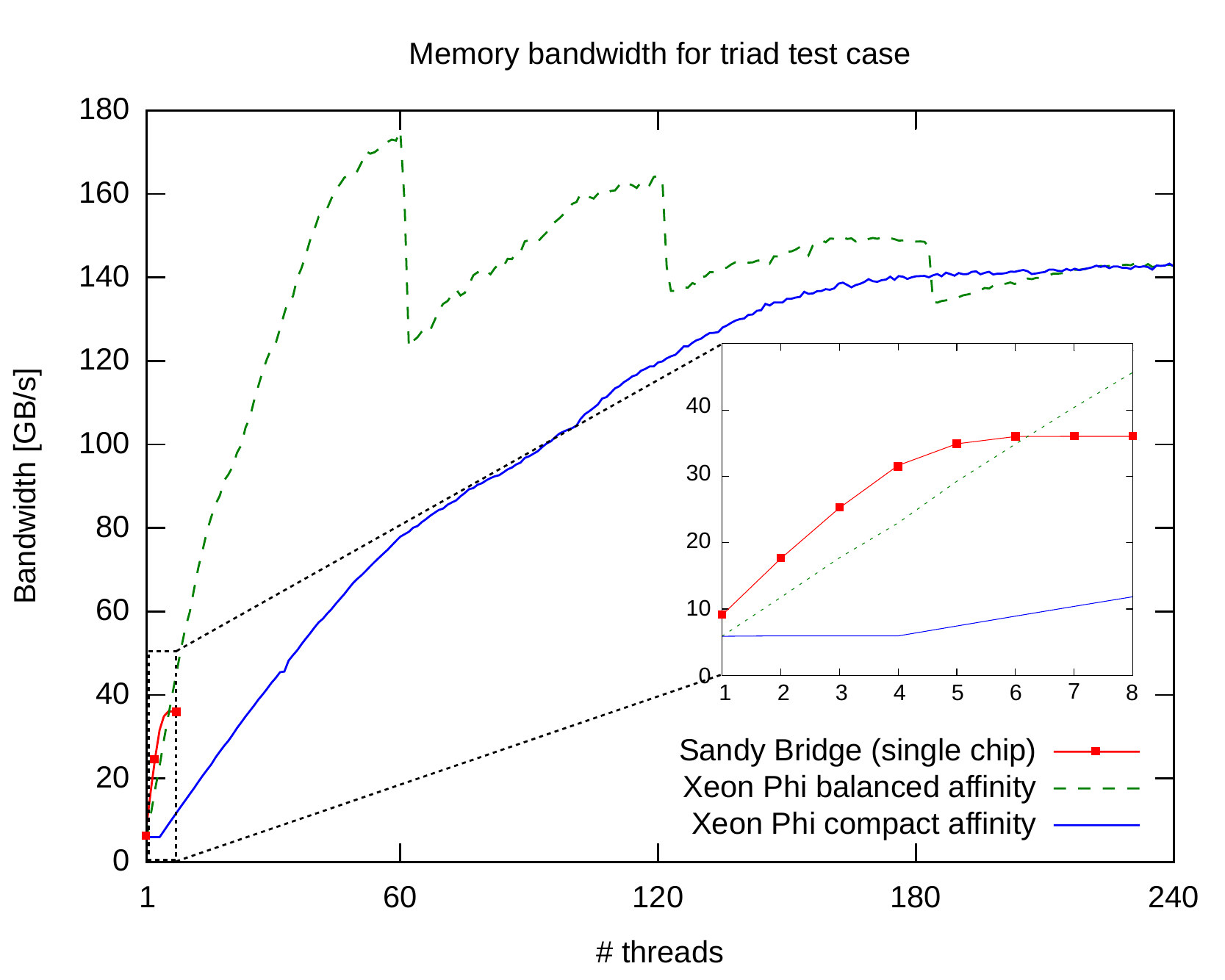}  
\vspace*{-0.3cm}
  \caption{Memory bandwidth measured by the \stream benchmark (triad case) on \intel{} Sandy Bridge and \intel{} Xeon Phi.}
  \label{fig:phi_mem_bw}
\end{figure}

Although the peak performance can be reached by some specific computing kernels, there is no way the peak memory bandwidth given by the vendor can be reached in practice.
The reasons for this are deeply buried inside the computing core and how the different memory cache hierarchies including the registers are handled.
However, the sustainable memory bandwidth can be evaluated by running a micro-benchmark.
The results shown in Figure~\ref{fig:phi_mem_bw} have been obtained by running the STREAM benchmark\footnote{http://www.cs.virginia.edu/stream/} which is nowadays the standard benchmark to evaluate the sustained memory bandwidth.

One can notice that the peak sustainable memory bandwidth is not achievable in every condition.
On a single Sandy Bridge socket, at least five cores should be used to reach it. 
During the various experiments performed on a single Sandy Bridge processor, we
observed that the way threads are pinned inside the processor has no impact on the memory bandwidth.
On the Xeon Phi, on the other hand, the thread pinning or thread affinity plays a significant role.
We have seen in section~\ref{sec:phi_archi} that each core on the Xeon Phi supports four hardware thread units.
The \intel{} specific environment variable KMP\_AFFINITY is used to place the different threads explicitly on the different thread units.
One could have used the more standardized and more recent thread pinning functionalities of OpenMP 4.0 but we did not investigate this as the usage of KMP\_AFFINITY is comfortable on the MIC architecture.
The "compact affinity" fills the first core with four threads and then the second core with four threads and so on.
This means that with eight threads on the Xeon Phi with "compact affinity", only two physical cores out of the 60 are in use.
On the contrary, the "balanced affinity" distributes the threads as evenly as possible among the 60 cores.
With 60 threads with the balanced affinity, one thread is running on each core and, in this configuration, the maximum memory bandwidth of 174 GB/s is reached when the \stream benchmark is compiled with the appropriate flags to parameterize the prefetching\footnote{https://software.intel.com/sites/default/files/article/370379/streamtriad-xeonphi-3.pdf} and for a specific number of threads: 60 in our case.

\subsection{Latency micro-benchmarks}
\begin{figure}[htb]
  
  \includegraphics[width=\linewidth]{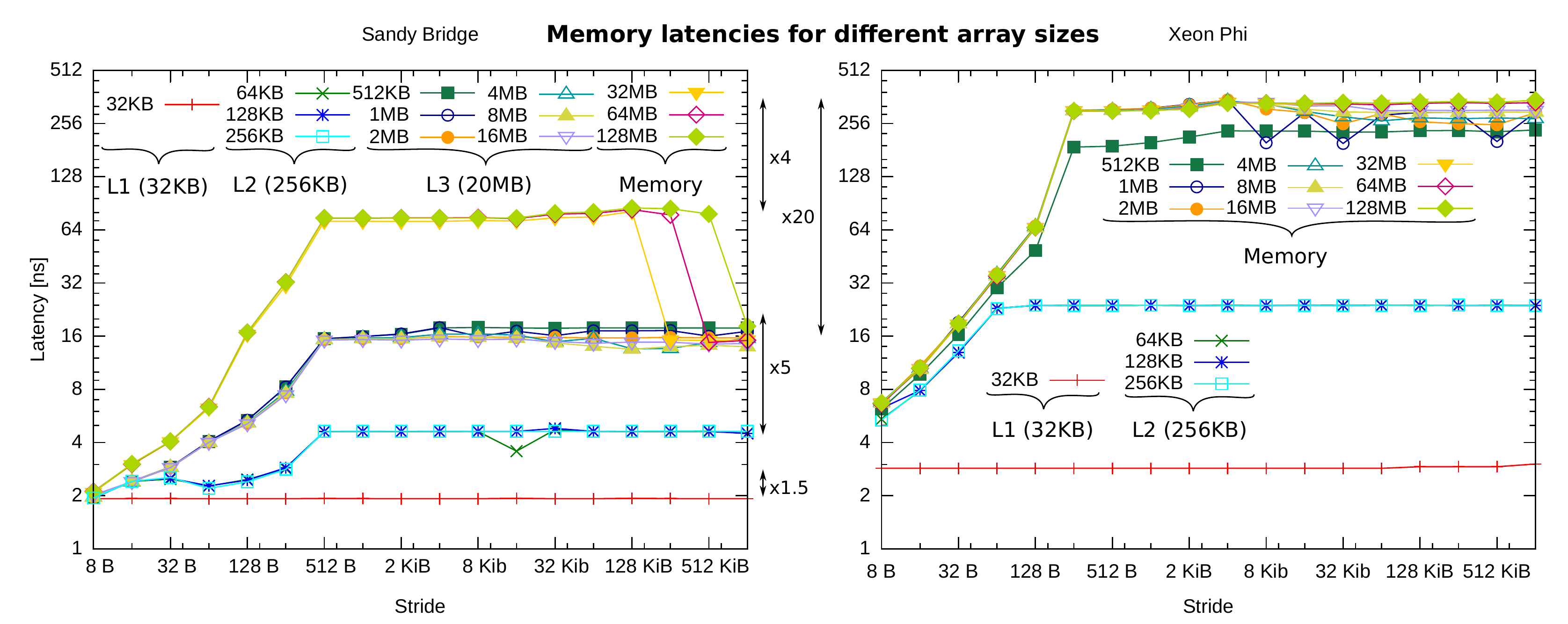}  
  \caption{Memory latencies measured by a pointer chasing benchmark on \intel{} Sandy Bridge and \intel{} Xeon Phi.}
  \label{fig:phi_mem_lat}
\end{figure}

The \stream benchmark used previously to measure the memory bandwidth accesses successively contiguous data elements.
Thanks to such regular pattern, data prefetchers can trigger the transfer of data from memory to the cache in advance so that the data is already in the cache when needed.
In the case of non-regular memory accesses, data prefetchers cannot help.
The memory access triggers a cache miss and the needed data has to be brought into the cache at that moment.
The time it takes to satisfy such memory access is called the memory latency and depends on the memory hierarchy in which the required data is residing at the time it is requested.
Figure~\ref{fig:phi_mem_lat} shows the results of a pointer chasing benchmark run on Sandy Bridge and Xeon Phi.
For a given array size, a pointer in memory accesses successively data from this array separated by a given stride.
This procedure is repeated a couple of thousands of times and the overall elapsed time is measured.
By dividing the elapsed time by the number of repetitions, we obtain the average latency of an operation for a given stride and array size.
When the full array fits into a memory level, the array is loaded once from memory into this cache hierarchy and is always accessed from there afterwards.
So the latency of the different memory levels can be observed on both Sandy Bridge and Xeon Phi by varying the array size.
One can see that the latencies are at least a factor four larger on Xeon Phi compared to Sandy Bridge as soon as the data is out of the L1 cache.
The worst case happens for data size comprised between 512KB and 16MB.
Indeed, they reside in L3 on Sandy Bridge with a latency of 16 ns whereas they reside in memory (i.e. not in cache) on the Xeon Phi with a latency of 330 ns.
This very strong effect is slightly smoothed out by the usage of the L2 cache of other cores.
If one compares the case where all data are in the memory with the case where all data are on other L2 caches, an improvement of 30\% on the latencies can be observed, which can be considered as a small impact. 
The results obtained on the Xeon Phi match the ones found in~\cite{fang:an}.

 These high latencies in combination with smaller cache sizes bring us to the conclusion that efficient cache reuse strategies are much more difficult to implement on the Xeon Phi than on the Sandy Bridge.
 Conversely, streaming implementations that access data as contiguous as possible enable the prefetchers to kick in and allow the application to benefit from the large memory bandwidth.
 But this does not come always automatically and fine-tuning of the prefetch mechanism at compile time or in the code with the help of pragmas are still of great importance on Xeon Phi to achieve such memory bandwidth.

 \subsection{Network micro-benchmark}
 \label{sub:network_micro_benchmark}

Within the standard partition of Helios, when two MPI tasks on different nodes communicate, the data transfer involves potentially the Infiniband (IB) port with the attached PCIe link and the QPI interface.
Even if this network structure is not completely homogeneous, considering it homogeneous is still a good approximation.
As the Xeon Phi is connected to the PCIe bus on the host, one additional communication link has to be potentially crossed and the assumption of a homogeneous network no longer stands.
For example, within a node of Figure~\ref{fig:phi_node_arch}, when a connection is established between CPU0 and CPU1, the data only has to cross the QPI interconnect whereas for a connection between MIC0 and MIC1, it has to cross PCIe, QPI and PCIe interconnects.
The worst case scenario happens when MIC0 on host0 communicates with MIC0 on host1 when only a single IB port is available on the node.
In this case, a connection has to cross successively PCIe, QPI, PCIe, IB, PCIe, QPI and PCIe interconnects. 

We have performed our own measurements of the network performance on three different MIC clusters, namely Robin located in Grenoble, France at the Bull R\&D center, Supermic located in Garching, Germany at LRZ and Eurora located in Bologna, Italy at Cineca.
On Robin a single IB port is connected to CPU1.
In addition, MIC0 is connected to CPU0 and MIC1 is connected to CPU1.
The performance is measured by the ping pong test of the \intel{} MPI Benchmark suite.
We define the latency as the time taken by a one byte message to travel from one task to the other and we define the bandwidth as the bandwidth achieved while transferring a $4.2$MB message between two different tasks.
Only the results on Robin are shown in Table~\ref{table:res_net} as the ones on Supermic and Eurora are similar. 
  
\begin{figure}
\centering
\begin{subfigure}[t]{0.5\textwidth}
\centering
\begin{tabular}{|c|c|c|c|c|}
  \hline
  \multirow{3}{*}{Host0}
 & CPU1  & 0.69 &  &   \\
 & MIC0  & 4.90 & 2.73 &   \\
 & MIC1  & 4.31 & 7.56 & 3.12  \\
  \hline
  \multirow{3}{*}{Host1}
 & CPU1  & 2.20 &  &   \\
 & MIC0  & 4.71 & 9.04 &   \\
 & MIC1  & 4.66 & 7.93 & 6.92  \\ 
  \hline
 \multicolumn{2}{c|}{} & CPU1 & MIC0 & MIC1\\
  \cline{3-5}
 \multicolumn{2}{c|}{} & \multicolumn{3}{|c|}{Host0}  \\
  \cline{3-5}
\end{tabular}
\caption{Latencies ($\mu$s)}
\label{table:res_net:lat}
\end{subfigure}%
~ 
\begin{subfigure}[t]{0.5\textwidth}
\centering
\begin{tabular}{|c|c|c|c|c|}
  \hline
  \multirow{3}{*}{Host0}
& CPU1  & 5029\,$^*$  &  &   \\
 & MIC0  & 456  & 2016 &   \\
 & MIC1  & 1609 & 416  & 2004  \\ 
  \hline
  \multirow{3}{*}{Host1}
 & CPU1  & 5729 &  &   \\
 & MIC0  & 418  & 273 &   \\
 & MIC1  & 1608 & 418 & 969  \\ 
  \hline
 \multicolumn{2}{c|}{} & CPU1 & MIC0 & MIC1\\
  \cline{3-5}
 \multicolumn{2}{c|}{} & \multicolumn{3}{|c|}{Host0}  \\
  \cline{3-5}
\end{tabular}
\caption{Bandwidth (MB/s)}
\label{table:res_net:bw}
\end{subfigure}

\caption{Network performance between different computing elements of the Robin cluster.
Table~\ref{table:res_net:lat} present the measured latency and Table~\ref{table:res_net:bw}, the bandwidth.
The upper part presents intra-node performance whereas the bottom presents inter-node performance. \\$^*$ {\small This intra node bandwidth smaller than the inter node one is somehow strange but reproduceable on several MIC clusters (we did not investigate any further).}}
  \label{table:res_net}
\end{figure}

Considering intra-node latencies shown in Table~\ref{table:res_net:lat}, one can notice a factor four difference between the latency observed by two tasks running on the same CPU and two tasks running on the same MIC.
The latency further degrades for the two possible pairs CPU-MIC and is finally almost 11 times worse for a MIC0-MIC1 connection than a CPU-CPU connection.
For inter-nodes connections between Host0 and Host1, latencies for MIC-MIC connections further degrade but not that much compared to the degradation of the CPU-CPU latency.
Surprisingly, the latency of the CPU-MIC0, CPU-MIC1 and MIC0-MIC1 connections does not degrade that much when crossing the node boundary.
The PCIe links between the CPUs and the MIC seem to be the larger bottleneck to these latencies.  
Considering intra-node bandwidth (upper right table), one can notice similarly a factor of more than two between the bandwidth measured between two tasks running on the same CPU and two tasks running on the same MIC.
Performance degrades for CPU-MIC1 connections and up to a factor of ten for CPU-MIC0 or MIC1-MIC0 connections.
Going across nodes further degrades performance, especially for MIC-MIC connections and down to 273 MB/s for a MIC0-MIC0 connection.

The results obtained confirm the ones shown in Figure~9 of~\cite{Karpusenko}.
The authors ran their benchmarks on two nodes, each equipped with only a single IB port and 4 MIC cards.
The worst case scenario being a MIC0-MIC0 connection in our case corresponds to MIC3-MIC3 connection in their case.
Fortunately, the hardware of the MIC partition on Helios is more homogeneous: each node has two IB ports, one connected to each socket.
This means that a MIC0-MIC0 connection in Figure~\ref{fig:phi_node_arch} will go through PCIe, PCIe, IB, PCIe and PCIe, exactly like a MIC1-MIC1 connection.
But having the hardware does not seem to be enough.
Indeed, Supermic is equipped with it but still demonstrates similar performance as Robin which has only one IB port connected to CPU1.

In short, the network performance is very inhomogeneous between the different computing elements of such MIC cluster.
If one considers the native mode use case where a standard MPI application is deployed, the communication performance will likely be limited by the slowest path.
So one should expect latencies five times larger and bandwidth 20 times lower than on the network of a "standard cluster" connecting only CPUs.

The reasons for this network performance inhomogeneity are not yet completely understood.
Setting the appropriate software stack and operating system configurations for that type of hardware seem to be complex and still an issue.
As a consequence, we have decided to keep inter- and intra-node aspects for a later version of the MIC processor family and instead focus on the intra-MIC optimizations in the remaining of this paper.



\section{\label{framework}Application framework}

\subsection{\gysela{} application}
The \gysela{} code is a non-linear 5D global gyrokinetic full-f code
which performs flux-driven simulations of ion temperature gradient
driven turbulence (ITG) in the electrostatic limit with adiabatic
electrons. It solves the standard gyrokinetic equation for the full-f
distribution function, \textit{i.e.} no assumption on scale separation
between equilibrium and perturbations is done. This 5D equation is
self-consistently coupled to a 3D quasineutrality equation. The code
also includes other features not described here (ion-ion collisions,
several kind of heat sources). The code has the originality to be
based on a semi-Lagrangian scheme\,\cite{son99} and it is parallelized using an
hybrid OpenMP/MPI paradigm\,\cite{gl07,crous09}. 

The coordinate systems we consider is as follows. The spatial
coordinates consist in $(r,\theta)$ the polar coordinates in the
poloidal plane (the origin for the radius $r$ is the magnetic axis and
$\theta$ is the angle), $\varphi$ the angle in the toroidal
direction. The velocity parallel to the magnetic field is $\vpar$. The
magnetic moment \mbox{$\mu=m v_\perp^2/(2B)$} is an adiabatic
invariant with $v_\perp$ the velocity in the plane orthogonal to the
magnetic field. The computational domain is defined on
$r\!\in\![\rmin,\rmax],\theta\!\in\![0,2\pi],\varphi\!\in\![0,2\pi],\vpar\!\in\![v_{\min},v_{\max}],\mu\!\in\![\mu_{\min},\mu_{\max}]$.

Let $\vec{z}=(r,\theta,\varphi,\vpar,\mu)$ be a variable describing
the 5D phase space. The ionic distribution
function (main unknown) of the guiding-center is $\fbar(\vec{z})$. The time evolution of this quantity  is
governed by the gyrokinetic Vlasov equation:
\begin{equation}
  \partial_t\fbar + \frac{1}{\Bstar}\nabla_{\vec{z}}\cdot\left(\frac{{\rm d}\vec{z}}{{\rm d}t}\Bstar\fbar\right) = 0\ .
\end{equation}


The $\Bstar$ and $\vecbstar$ terms are defined as:
\begin{equation*}
  \Bstar = B+\frac{m\vpar}{e\,B}\mu_0\vecb\cdot\vec{J},\ \ \ \ \ \bstar = \frac{\vec{B}}{\Bstar}+\frac{m\vpar}{e\Bstar}\,\frac{\mu_0\vec{J}}{B}\ . \label{vecbstar}
\end{equation*}
The magnetic field is $\vec{B}$ (with $B$ the magnitude of
$\vec{B}$), $\vec{J}$ stands for the plasma current density. Vacuum
permittivity is denoted $\mu_0$.  The particle motion described by the
previous Vlasov/transport equation is coupled to a 3D quasi neutral solver
(Poisson-like solver) that computes the electric potential
$\footnotesize\phi(r,\theta,\varphi)$:
\begin{equation*}
\frac{e}{T_e}(\phi-{\langle\phi\rangle)} = \frac{1}{n_0}\int{}J_0(\bar{f} - \bar{f}_{init})\dv + \rho_i^2\nabla_\perp^2 \frac{e\phi}{T_i}\ .
\end{equation*}\\[-1mm]

We will not describe this last equation, but details can be found in
\cite{lin,virginie,RR-7595}. This Poisson-like equation gives the
electric field $\phi$ that corresponds to the particle distribution $\fbar$ at
each time step $t$. Then, this electric field produces a feedback in the
Vlasov equation through the term $\frac{{\rm d}\vec{z}}{{\rm
    d}t}\Bstar\fbar$ (details can be found for example in
\cite{gl14}). Practically, at each time step both Vlasov and Poisson
equations are solved successively. We will investigate the
Vlasov solver  in this paper, which  represents the most costly  part (more than
98\% of a sequential run if diagnostics and outputs are deactivated).

We have chosen to present here configurations with only a single
$\mu=0$ value. It means, we consider the 4D drift-kinetic model which
is the backbone of the 5D gyrokinetic models and relevant to build
numerical schemes.  To simplify the analysis and avoid possible
problems due to MPI communication performance
(\tit{cf.}~subsection~\ref{sub:network_micro_benchmark}), we will
consider in the following a mini-application that does not use MPI and
is only parallelized using the OpenMP paradigm. All floating point computations of the codes presented in this paper use
double precision numbers~(64 bits).

\subsection{Splitting scheme versus 4D advection, benefits expected}
\label{split_scheme}

The usual way to perform a single \textit{advection} (transport mechanism caused
by the flow) in the \gysela{} code\,\cite{virginie,gl07}
consists of a series of directional advections:
${\textstyle (\hat{\vpar}/2,\hat{\varphi}/2,\hat{r\theta},\hat{\varphi}/2,\hat{\vpar}/2)}$. Each
directional advection is performed with the semi-Lagrangian
scheme. This Strang-splitting converges in ${O({{\Delta{}t}^2})}$. It
decomposes one time step into four 1D advections and one central 
2D advection (in the poloidal plane $(r,\theta$)). 
Algorithm \ref{algo1D} shows a single  1D advection in the $\varphi$ direction.\\
\begin{algorithm}[th]
\vspace*{.3cm}
\SetLine
\For{All grid points $(r_i,\theta_j,{\vpar\,_l})$}{
  $\eta(\varphi=*) \leftarrow{}$ spline coefficients computed from the 1D function $f^n(r_i,\theta_j,\varphi=*,{\vpar\,_l})$\; 
  \For{All grid points $\varphi_k$}{
    $(\varphi_k)^\star \leftarrow{}$ foot of characteristic for one time step $\Delta{}t$ that ends at $(r_i,\theta_j,\varphi_k,{\vpar\,_l})$\;
    Interpolate $f^n$ at location $(\varphi_k)^\star$ using $\eta$ coeff.\; 
    $f^{n+1}(r_i,\theta_j,\varphi_k,{\vpar\,_l}) \leftarrow {\rm the\ interpolated\ value}$\;
  }
}
\caption{1D advection with Semi-Lagrangian scheme}
\label{algo1D}
\end{algorithm}
Let us now consider an alternative method that avoids the Strang
splitting. Algorithm \ref{algo4D} sketches the corresponding 4D
numerical scheme (\textit{Nosplit} case, see also \cite{gl14}). The difference with the previous
scheme is twofold. The feet of the characteristics has to be computed
in four dimensions and the spline coefficients come from a tensor
product of splines in four dimensions. The OpenMP
parallelization and SIMD vectorization of the whole Algorithm \ref{algo4D} represent the challenge
to tackle (especially the 4D interpolations~-~line 5).\\
\begin{algorithm}[th]
\vspace*{.3cm}
\SetLine
$\eta(r=*,\theta=*,\varphi=*,\vpar=*) \leftarrow{}$ compute spline coeff. from the \\
\Indp\Indp  4D function $f^n(r=*,\theta=*,\varphi=*,\vpar=*)$\; \Indm\Indm \linesnotnumbered
\For{All grid points $(r_i,\theta_j,\varphi_k,{\vpar\,_l})$}{
  $(r_i,\theta_j,\varphi_k,{\vpar\,_l})^\star \leftarrow{}$ foot of characteristic that ends at $(r_i,\theta_j,\varphi_k,{\vpar\,_l})$ \; \linesnotnumbered
  Interpolate $f^n$ at location $(r_i,\theta_j,\varphi_k,{\vpar\,_l})^\star$ using $\eta$ coeff.\; 
  $f^{n+1}(r_i,\theta_j,\varphi_k,{\vpar\,_l}) \leftarrow {\rm the\ interpolated\ value}$\;
}
\caption{4D advection scheme  with Semi-Lagrangian scheme}
\label{algo4D}
\end{algorithm}
Let us have a look to the algorithmic complexity of the two
approaches:  splitting  scheme  (denoted  \textit{Split}), 4D  advection  scheme
(denoted \textit{Nosplit}). To
oversimplify this short analysis, we will focus on the interpolation
operator, and the spline coefficients computation. Cubic spline interpolation is chosen as it has been shown
that such Vlasov solver is well resolved with such interpolation
operator. We leave aside the cost due
to the computation of the feet of the characteristics. Furthermore, we assume that we have a constant displacement
field for the advection (no computation required). Let us denote
$N_{all}$ the number of points in the domain, \tit{i.e.}
$N_{all}=N_r\,N_\theta\,N_{\vpar}\,N_\varphi$.


For the \textit{Split} case, each 1D advection requires: $7\,N_{all}$ floating
point operations (it means 3 additions and 4 multiplications per grid
point: see the stencil in Figure~\ref{lag1d_dir}, p. \pageref{lag1d_dir}) and $N_{all}$ read memory accesses, and the same amount of
write  memory   accesses.   The 1D spline  coefficients  derivation   (line  2  of
Algo. \ref{algo1D}) costs $10\,N_{all}$ FLOP
(a solver performs a single sweep down and up using a small LU decomposition).  We assume
that each advection is performed with a \textit{very good} locality in cache
memory that prevents from loading the same memory
reference  two   times  \textit{during  a   single  advection},  especially   the  spline
coefficients remain in the cache. The 2D advection implies: $35\,N_{all}$ FLOP (see the stencil in Figure~\ref{lag2d_dir}, p. \pageref{lag2d_dir}),
$N_{all}$ read memory accesses, and the same amount of write memory
accesses. The 2D spline coefficients derivation leads to $10\,N_{all}$ FLOP.
To summarize, the splitting scheme, meaning 5 advections, has the following cost: $113\,N_{all}$ FLOP, $5\,N_{all}$ in both read and write memory accesses.

Concerning the \textit{NoSplit} case, the computational cost of the 4D interpolator
is quite high:  $595\,N_{all}$ FLOP. The spline coefficients  derivation using a
tensor product in four dimensions leads to $20\,N_{all}$ FLOP.
Considering memory accesses, the
distribution function $f^n$ is accessed one time to compute the spline
coefficients.  The  spline coefficients  are  first written  (lines~1  and~2  of
Algo. \ref{algo4D}) and then read in order to compute the interpolations (line~5). Finally, the interpolation
result is written (line 6). So in total, the algorithm performs $2\,N_{all}$
read  memory  accesses  and   $2\,N_{all}$  write  memory  accesses,  and  means
$615\,N_{all}$ FLOP.

If one compares both methods, the \textit{Nosplit} case computes 6 times more FLOP
but performs less memory accesses. In first approximation, the 1D and
2D advections are mainly memory-bound kernels (as we will see
afterwards), whereas the 4D approach is clearly CPU-bound. For
computing units that can perform a very large number of FLOP per
transferred byte (as the Xeon Phi is), the 4D approach is
better suited than the splitting approach.
In the following, we will mainly target the 4D algorithm for the porting on Xeon Phi.

\subsection{\gysela{} mini-app}
\label{firstminiapp}

In practice, we have used the \texttt{OpenMP} paradigm on a shared memory
node throughout all the codes presented hereafter. In this study, the
focus was made on the optimization/vectorization issue inside one
single node (2 sockets, Sandy Bridge, 16 cores) and one Xeon Phi (60
cores). This enables to look at optimization and vectorization issues,
and compare raw performances. The \gysela{} mini-app is a 4D drift-kinetic code
($\mu=0$) in which we have removed a lot of diagnostics and physics
submodels from the original \gysela{} code. Furthermore, the mini-app
contains the 4D advection capability which is not yet used in the
production code. A small run with this mini-app on one single Sandy Bridge node 
takes 10 hours for the following domain size:
$N_r=128,N_\theta=256,N_\varphi=32,N_{\vpar}=64$ with 4000 time steps. This run fully used 16 cores thanks to \texttt{OpenMP} parallelization. The code was compiled with \texttt{-xAVX} compiler flag in order to have some automatic vectorization whenever it is possible.

A profiling of the application using a single process with 16 threads has been performed on the Sandy
Bridge partition with the Scalasca toolkit. As a result 98\% of the run time is concentrated in one single routine and can be split into:
\begin{itemize}
  \item Spline construction 10\%
  \item Vlasov Solver (4D advection) 89\% split into:
     \begin{itemize}
       \item Computation of the feet of the characteristics 41\%
       \item Interpolations with 4D stencil using spline coefficients and feet 59\%
     \end{itemize}
\end{itemize}

The spline construction represents 10\% of the execution time which is
mostly spent in memory copies.  In principle, it shows some space for
improvements as it has not been completely parallelized.  However, as
it represents only 10\%, we will focus on the parallel region which
consumes 89\% of the dominant subroutine and mainly on the 4D
interpolation routine (51\% of the total run time).

We first tried a \textit{top-down} approach to improve the
performance of the 4D interpolation kernel. We extracted the most
computation intensive part of the kernel in a single subroutine of less than 200
lines of code. Then, a lot of optimization techniques were investigated
on this subroutine to get a vectorization of parts of the kernel and cache-friendly behaviour. We have speed up the initial version by a factor 2 on medium test cases on the Xeon Phi (domain size $128\times{}128\times{}64\times{}32$, 45s for one advection step with the initial version and 25s after the improvements have been done).
Nevertheless, on a Sandy node with 16 cores and a similar test case, the code is eight times \textit{faster} (3.3s per advection step) than on one Xeon Phi device (25s). Even though we have explored many techniques, we failed to reach the level of performance one can expect from Xeon Phi (\textit{i.e.} Sandy should have been two times \textit{slower} than Phi on well optimized/vectorized code). 

Therefore, we switched to a different approach further discussed in the following sections. In this \textit{bottom-up} approach, we start with the study of small kernels that can reach high performance on Phi. And then, from this experience we build up more complex kernels while trying to keep the same level of performance.


\section{\label{kernels}Benchmarking interpolation kernels}
\subsection{Choosing interpolation kernels}

The choice of interpolation method in a semi-Lagrangian scheme is
crucial. It determines the numerical quality of the scheme and its
computational cost. The tensor product is employed in this
work to achieve multi-dimensional interpolations. Therefore, we 
need to fix first the method for the 1D interpolation. We will
investigate the cubic spline scheme, which is known to be a
well-balanced compromise between cost and quality for plasma
simulations. This is the scheme used in the \gysela{} code in
production. Nevertheless, cubic splines lead to rather complex computational
kernel. Thus,
we have chosen to first look at Lagrange polynomials\footnote{The
numerical quality of Lagrange polynomials is not as good as
splines in our context\cite{fil01}.}. The Lagrange interpolator is simpler and thus
gives us the opportunity to easily investigate several
optimization alternatives. In addition the $4^{th}$ order Lagrange polynomial is close to
cubic spline in terms of computational cost (except that we do not need
to compute spline coefficients) and in terms of data access
pattern. That will give us hints to optimize cubic splines afterwards.

We will focus on 4 kernels based on Lagrange polynomials of order 4
using tensor product in 1, 2, 3 and 4 dimensions. Then, we will tackle more complex kernels, 2D cubic spline
interpolation and finally 4D spline interpolation. In order to have
small codes which can be easily handled and tuned, we will define in
the next subsection a small model (designed for verification purposes)
corresponding to a very simplified Vlasov equation.

\subsection{Synthetic model - setup}

We intend to evaluate the interpolation kernels within the
Semi-Lagrangian scheme. In order to address this issue, we define a much simpler 
equation than the gyrokinetic setting described earlier as well as
an analytical test case.  The advection equation that we consider is
the following (with $D\in[1..4]$ depending on the kernel under evaluation): \be
\label{eqadvcst}
\partial_t f + v\,\partial_x f = 0, \;\; x\in [0, 2\pi]^D, \;\; t\geq 0\,.\\
\ee
\noindent We assume  a constant velocity field\hspace*{.5cm} 
$\textstyle v=
\begin{pmatrix}
v_1 & \hdots & v_D
\end{pmatrix}^T
$.\\ With an initial state\hspace*{.15cm}
$
f(t\!=\!0)=\Pi_{d=1..D}\ sin(m_d\,x_d)
$, where $m_d$ are positive integers,
the analytical solution at time $t$ is\hspace*{.15cm}
$f_{analytic}(t)=\Pi_{d=1..D}\ sin(m_d\,(x_d-v_d\,t))$. This known
solution is helpful to verify the implementation of the scheme solving
Eq. \ref{eqadvcst} numerically. For a constant velocity field, the
computation of the feet of characteristics can be done once for all,
and is cheap to perform. Therefore, the numerical solver execution
time is dominated by the computations due to the interpolation operator
that represents more than 90\% of total execution time.

\subsection{Code description}

To build up a code and in order to optimize it on Xeon Phi, there are a few
works in the literature that can help. Here is a
short list of some papers and
documentations\,\cite{Gepner14,phi_inst_set,rahman,fang:an,Ros13}.

The simplest interpolation kernel is shown on
Figure~\ref{lag1d_dir}.   The    function  \texttt{access\_f}   is   an
accessor\footnote{The accessor is useful to switch
between C and Fortran language easily, and also to insert padding in the
multi-dimensional arrays, to have aligned access, or to avoid cache trashing.}
to  get/set the value  of a  distribution function. The input
distribution function is \texttt{f0} (at time $t$), the output is
\texttt{f1} (time $t+\Delta{}t$). The variables \texttt{coeff[1-4]}
are set depending on the velocity field. For
the sake of simplicity, the velocity is assumed to be small (or
$\Delta{}t$ small) in order to have a foot of characteristic in the
left or right cell near the departure point $(x_1,x_2,x_3,x_4)$.
In this peculiar situation, one
does not need to establish which grid point is the closest to the foot
of the characteristic as the characteristic origin is this point. This simplifies
the code  and allows to isolate  the code that performs  the interpolation. This
unrealistic assumption will be relaxed in the last section of this paper.\\
The kernel in Figure~\ref{lag1d_int}
performs the same calculation as Figure~\ref{lag1d_dir} but using dedicated intrinsics\,\cite{phi_inst_set}, that
enables to use Xeon Phi SIMD instruction set.

\begin{figure}[h!]
\vspace*{-0.2cm}
\begin{minipage}[b]{8cm}
\scalebox{.55}{
\begin{minipage}{13cm}
\begin{lstlisting}[language=C, firstline=1, lastline=14]
#pragma omp parallel for collapse(3)
 for (x1=0; x1<Nx1; x1++) {
  for (x2=0; x2<Nx2; x2++) {
   for (x3=0; x3<Nx3; x3++) {
#pragma vector nontemporal (f1)
#pragma vector always  
    for (x4=0; x4<Nx4; x4++) {
     access_f(f1,x4,x3,x2,x1) = // OUTPUT data f1
       coef1 * access_f(f0,x4-1,x3,x2,x1) + // INPUT data f0
       coef2 * access_f(f0,x4  ,x3,x2,x1) + 
       coef3 * access_f(f0,x4+1,x3,x2,x1) + 
       coef4 * access_f(f0,x4+2,x3,x2,x1);  
    } } } }
\end{lstlisting}
\end{minipage}
}
\caption{Lagrange 1D code - with directives}
\label{lag1d_dir}
\end{minipage}
\begin{minipage}[b]{8cm}
\scalebox{.55}{
\begin{minipage}{13cm}
\begin{lstlisting}[language=C, firstline=16, lastline=36]
#pragma omp parallel for collapse(3)
 for (x1=0; x1<Nx1; x1++) {
  for (x2=0; x2<Nx2; x2++) {
   for (x3=0; x3<Nx3; x3++) {
    for (x4=0; x4<Nx4; x4+=8) {
      ptread = &(access_f(f0,x4,x3,x2,x1));
      // read input data
      tmpr2 = _mm512_load_pd  (ptread);
      tmpr1 = _mm512_loadunpacklo_pd(tmpr1, ptread-1);
      tmpr1 = _mm512_loadunpackhi_pd(tmpr1, ptread-1+8);
      tmpr3 = _mm512_loadunpacklo_pd(tmpr3, ptread+1);
      tmpr3 = _mm512_loadunpackhi_pd(tmpr3, ptread+1+8);
      tmpr4 = _mm512_loadunpacklo_pd(tmpr4, ptread+2);
      tmpr4 = _mm512_loadunpackhi_pd(tmpr4, ptread+2+8);
      // 1+2+2+2=7 flop per loop iteration
      tmpw = _mm512_mul_pd(tmpr1,coeff1);
      tmpw = _mm512_fmadd_pd(tmpr2,coeff2,tmpw);
      tmpw = _mm512_fmadd_pd(tmpr3,coeff3,tmpw);
      tmpw = _mm512_fmadd_pd(tmpr4,coeff4,tmpw);
      // write output data
     _mm512_store_pd (&(access_f(f1,x4,x3,x2,x1)), tmpw);
\end{lstlisting}
\end{minipage}
}
\caption{Lagrange 1D code - with intrinsics}
\label{lag1d_int}
\end{minipage}
\end{figure}
Figure~\ref{lag2d_dir} shows the interpolation with the 2D tensorial
product of Lagrange polynomial (order 4). The number of FLOP grows
from 7 in the inner loop of Figure~\ref{lag1d_dir} to 35 in
Figure~\ref{lag2d_dir}.
\begin{figure}[h!]
\begin{minipage}[b]{8cm}
\scalebox{.55}{
\begin{minipage}{13cm}
\begin{lstlisting}[language=C, firstline=1, lastline=25]
#pragma omp parallel for collapse(3)
 for (x1=0; x1<Nx1; x1++) {
  for (x2=0; x2<Nx2; x2++) {
   for (x3=0; x3<Nx3; x3++) {
#pragma vector nontemporal (f1)
#pragma vector always  
    for (x4=0; x4<Nx4; x4++) {
     access_f(f1,x4,x3,x2,x1) = 
       coefb1 * (coefa1 * access_f(f0,x4-1,x3-1,x2,x1) + 
		 coefa2 * access_f(f0,x4  ,x3-1,x2,x1) + 
		 coefa3 * access_f(f0,x4+1,x3-1,x2,x1) + 
		 coefa4 * access_f(f0,x4+2,x3-1,x2,x1)  ) +  
       coefb2 * (coefa1 * access_f(f0,x4-1,x3  ,x2,x1) + 
		 coefa2 * access_f(f0,x4  ,x3  ,x2,x1) + 
		 coefa3 * access_f(f0,x4+1,x3  ,x2,x1) + 
		 coefa4 * access_f(f0,x4+2,x3  ,x2,x1)  )  +
       coefb3 * (coefa1 * access_f(f0,x4-1,x3+1,x2,x1) + 
		 coefa2 * access_f(f0,x4  ,x3+1,x2,x1) + 
		 coefa3 * access_f(f0,x4+1,x3+1,x2,x1) + 
		 coefa4 * access_f(f0,x4+2,x3+1,x2,x1)  )  +
       coefb4 * (coefa1 * access_f(f0,x4-1,x3+2,x2,x1) + 
		 coefa2 * access_f(f0,x4  ,x3+2,x2,x1) + 
		 coefa3 * access_f(f0,x4+1,x3+2,x2,x1) + 
		 coefa4 * access_f(f0,x4+2,x3+2,x2,x1)  )  
       } } } }
\end{lstlisting}
\end{minipage}
}
\caption{Lagrange 2D code - with directives}
\label{lag2d_dir}
\end{minipage}
\vspace*{-0.2cm}
\end{figure}

\subsection{Comparing Sandy Bridge versus Phi performance}
\label{kernel_sb_vs_phi}

In this subsection, we compare the performance of several kernels on a
Xeon Phi in native mode, against one Sandy Bridge node (two sockets,
16 cores). In theory, the Phi can outperform Sandy by a factor 3 on
compute-bound kernels (considering ratio of peak processing performance, the peak is established according to \intel{} specification sheet), and a factor 2
for memory-bound kernels (ratio of the \textsc{stream} benchmark
performance). Table \ref{perf_kernels} shows the obtained results. The compilation flags used on Sandy for the \intel{} C compiler are \texttt{-O3  -mmic -openmp}, whereas the flags for Phi are \texttt{-O3  -xAVX -openmp}.
\begin{table}[h!]
\medskip
\begin{tabular}{|ll|ll|ll|}
\hline
\multicolumn{2}{|c|}{Kernel/archi.} & \multicolumn{2}{c|}{Processing perf.} & \multicolumn{2}{c|}{Memory bandwidth}\\
\hline
Lagrange 1D & \textbf{Phi}  &  46 \gf{} &(5\% peak) & 106 \gbs{} & (81\% \textsc{stream})\\
            & \textbf{Sandy} &  25 \gf{} & (7\% peak) &  57 \gbs{} & (81\% \textsc{stream})\\
\hline
Lagrange 2D & \textbf{Phi} & 250 \gf{} & (25\% peak) & 111 \gbs{} & (85\% \textsc{stream})\\
            & \textbf{Sandy} & 134 \gf{} & (39\% peak) & 59 \gbs{} & (84\% \textsc{stream}) \\
\hline
Lagrange 3D & \textbf{Phi} & 228 \gf{} & (23\% peak) & 25 \gbs{} & (19\% \textsc{stream})\\
            & \textbf{Sandy} & 156 \gf{} & (46\% peak) & 17  \gbs{} & (25\% \textsc{stream}) \\
\hline
Lagrange 4D & \textbf{Phi} & 160 \gf{} & (16\% peak) & 4.3 \gbs{} & (3.3\% \textsc{stream})\\
            & \textbf{Sandy} & 145 \gf{} & (42\% peak)& 3.9 \gbs{} & (5.6\% \textsc{stream}) \\
\hline
\end{tabular}
\medskip
\caption{Performance of interpolation kernels}
\label{perf_kernels}

\vspace*{-0.6cm}
\end{table}
The first two kernels (Lagrange 1D and 2D) are dominated by the memory
bandwidth constraint. They achieve on both architectures more than 80\%
of the maximal bandwidth, which is very satisfactory. The measurement
of maximal bandwidth was established with \textsc{stream} benchmark -
\textsc{triad} operator and we measure 70 \gbs{} on Sandy, 130 \gbs{}
on Phi\footnote{Using  236 threads  on Phi which  is not  the best that  one can
  achieved ideally, but it  corresponds to the number of threads  we use for computation
  kernels, and it is thus more meaningful.  Then, this setting is different 
from the one shown in previous section \ref{sec:mic_membench}.}. Practically, one can thus achieve the expected speedup of Phi
over Sandy on a memory-bound kernel.

The 3D and 4D interpolation kernels are clearly compute-bound, they
have several hundreds of FLOP per grid points to perform.  The
Lagrange 3D implementation has good performance, but the Phi gets a 1.5
speedup over Sandy which is not the expected 2 or 3. The Lagrange 4D
kernel is not fast enough for the Phi implementation. The
performance is close to the Sandy one, while one can expect a good
speedup for this kind of kernel (Phi was designed to tackle computation intensive algorithms).
By replacing the complex accesses for memory reads by simpler but fake accesses, 
performance is improved a lot (results are wrong of course). 
This shows that a major problem comes from the 4D stencil that implies complex
memory access pattern.

\subsection{Problems encountered in the kernels, techniques used to solve them}

Several techniques have been tested to optimize and tune the
performance of the interpolation kernels and to achieve the results
obtained in Table~\ref{perf_kernels}. In the following section, we
briefly describe the different investigated approaches and their effect.

\subsubsection{Memory-bound kernels} 

For Lagrange 1D and 2D, the outer loops were parallelized through OpenMP directives and there is
no dependency between loop iterations. The most inner loop is 
vectorized using SIMD features. Several versions of the code have been written 
using \intel{} intrinsics for Xeon as well as standard C or Fortran code plus \intel{} SIMD directives.
The intrinsic versions require to write explicit low-level algorithms. This ends up
with a non-portable code (specific to Phi architecture), and also some troubles to maintain the code. In the case of
kernels instrumented  with directives, it is quite difficult to find the correct
location for directives, and also which directive to use. Indeed, in its report, the compiler can say that a
loop or a block is vectorized, even if the quality of the vectorization
will not give large benefits for example by masking all elements of the vector but one.  A look at the generated assembly
code\footnote{Example of compilation flags to get the assembly code: \texttt{ifort -fcode-asm
    -Facode.s -c code.c -o code.o}} helps a lot to control if the
compiler is doing a good choice of SIMD instructions for
vectorization. At the end of the day, both versions (directives and
intrinsics) give approximately the same performance and we have checked
that the most inner loop contains nearly the same floating-point SIMD
instructions.

For these memory-bound kernels, the prefetch mechanism is important. In
practice, it means loading data in advance  thanks to ad-hoc directives or
intrinsics. We have observed a 20\% reduction of execution
time by good choice of prefetch parameters. To determine the best
choice for the prefetch parameters, a script that scans an extensive
set of possible combinations has been made.

It is also interesting to tune the kernels in order to work on aligned
data. We have studied the memory access pattern in the inner loop to
avoid any cache trashing as it is also a key factor. These two
techniques enable to save 20\% execution time on 1D/2D kernels.

Finally on Phi, we can observe large variations of execution time from one
run to the other (10\% is common). Then, interpreting the impact of
any optimization often requires to launch a bunch of runs in order
to get good statistics.

\subsubsection{Compute-bound kernels} 

Cache blocking (loop tiling) is crucial to
save computation time. As caches are small on Phi, designing
algorithms that are cache friendly is quite a hard task. Nevertheless,
we manage to save execution time with proper loop tiling on the 3D kernels
(50\% gain).

We have compared C and Fortran kernels that are similar line-by-line
(with SIMD directives for the vectorization). The language has some
influence on the performance (at the percent level), but on some
kernels C is faster, on some others the Fortran version is faster.
So no clear trend could be identified here. If the language does not seem to
be an issue, the way 
algorithms are written triggers different kind of optimizations
performed by compiler. These have a large impact on 
execution time. For example, splitting the
body of a 10-40 lines loop into multiple loops of less than 10 lines of code
each (if it is semantically correct) can lead to effective
speedups. Indeed, the compiler seems much more efficient on small 
loop bodies and fusing loops is a much easier task for the compiler 
than splitting them. The observed effects are much stronger 
on Phi than on Sandy Bridge.

We have designed
several kernels for the 3D/4D interpolation with intrinsics and
compared them to versions with SIMD directives. In some settings, SIMD
intrinsics version is the fastest version. It is not the same
statement as for memory-bound kernels, but it makes sense as the
compute-bound kernels are very sensitive to the way computations are
optimized.

For these 3D and 4D kernels, we observe that the best performance is
achieved on Phi with 170 up to 240 threads. This very fine grain
parallelism has to be compared to the 16 threads that are sufficient
on Sandy (2 sockets). So there is an impressive factor 10 on the number of required threads. One has to
exhibit much more parallelism to get the performance out of the Phi
device. It also means that for some domain sizes or parameter sets, it
is difficult to get enough threads to get the best performance on the
Phi.

These rather simple compute-bound kernels are vectorized with difficulty
by the \intel{} compiler despite the usage of the available SIMD directives. 
A small modification in the code can
lead to bad vectorization at some location. When this kind of event
happens, the code can slow down by a factor of 4
suddenly. 

It is extremely difficult to identify this kind of vectorization problem (compiler does
not report it) and to set up a work around. As a practical method, we detect
such weak link by commenting a group of code lines and looking at the impact on performance.
Of course, the numerical results are wrong, but iterating this process on other groups
of lines enables to localize the weak link and sometimes to find a cure.

\subsubsection{Getting efficient kernels} 
A partial conclusion of this study is that achieving performance even
on a simple kernel is a challenge on Phi. Compared to the relative
easiness to get a reasonably efficient code on Sandy, the programmer
has a lot of constraints to fulfill on Phi. The developer has to
interact finely with the compiler and profiling tools to improve the
performance to an acceptable level.

If it is possible to take care of some of the important points on a
simplified kernel code, it can be very difficult to satisfy some
constraints throughout a long program. The non exhaustive list of
theses points could be: good vectorization, fine grain parallelism
(threads), data alignment, cache blocking (adapted to small
caches), prefetching, loop splitting.

Applying these hints in a long code, in which the computation
time is distributed over several routines, is not simple. These
optimization issues can pose a serious challenge to programmers that
want to fully use the hardware. Section \ref{miniapp_port} is dedicated
to this topic.


\section{\label{miniapp_port}Porting mini-app on Xeon Phi}
In the previous section, we have studied computation kernels closer and closer to the ones used in \gysela{} so as to identify guidelines and good practices to reach acceptable performance on the Xeon Phi.
This section builds up on this experience to target a more realistic kernel that uses a 4D tensor product of splines instead of Lagrange polynomials and a variable displacement field for the feet of characteristics.
In a first phase, we keep focusing on the interpolation part similarly to previous sections while in a second phase, we analyze the complete advection code.
To do so, we first consider the complete advection code but we comment out the parts we want to ignore for the first phase.
This approach enables us to easily extend the previous interpolation kernels and to integrate it without bothering with other issues coming from other parts of the application.
A performance evaluation in this configuration is given. Finally, we removed the comments added for the first analysis, and timings on the complete advection code are shown.

\subsection{Advection kernel design}

The kernel we design in this section is not a synthetic kernel anymore but is instead a complete implementation of Algorithm~\ref{algo4D} intended to be included in the \gysela{} code. As we have observed in previous paragraphs, there is only a small impact on performance if one choose SIMD directives instead of intrinsics for vectorization. For maintenance issue and because \gysela{} is a Fortran code, we made the choice to write the kernels in Fortran. 
Designing a complete kernel increases the amount of code that has to be optimized compared to the small (typically less than 150 lines of code) kernels of the previous sections.
This is the first source of increased complexity.

Another complexity comes from the use of a variable displacement field for the feet of the characteristics.
In the previous simple kernels we took purposely a very small displacement.
In this configuration, we do not need to recover the index of nearest grid point at the foot as it is the same as the grid point of the departure.
The fact that the foot can be everywhere in the computation domain must be considered for realistic cases (we will see this in section \ref{bench_true}).
From one given grid point $x$, the Semi-Lagrangian scheme uses the value at another location $x^{\star}$ that is possibly far away and has no predictable access pattern.
As a result, the data needed for the interpolation (spline coefficients) at the foot of the characteristic cannot be easily prefetched.
In addition, it generates an indirect access which induces a cost\footnote{Unit stride memory addressing is sadly not always achievable.}.
As expected due to the in-order architecture of the Phi, our measurements indicate that indirect accesses are relatively more costly on Phi compared to Sandy.

Also, this computation of the foot mixes floating-point (for the values) and integer (for the location) calculations that depend on each other. Furthermore, we did not found a way to vectorize both floating-point and integer operations efficiently. Then we focus on the vectorization of costly floating-point computations. This leeds to an innermost loop of the algorithm that is not completely vectorized with SIMD instructions since some integer operations remain. 

Taking all these aspects into account, we have applied strategies to reach good performance on the Xeon Phi.
This results in the code presented in Figure~\ref{protoadv_algo} for the outer loops and Figure~\ref{protoadv_internal} for the kernel called from these.
In order to come up with all these optimizations, we had to test several ways of writing the algorithms and closely look at the generated assembly code as it is always difficult to predict the impact of a change on the generated code.

\begin{figure}[h!]
\begin{minipage}[b]{8.5cm}
\scalebox{.55}{
\hspace*{1.2cm}
\begin{minipage}{12cm}
\begin{lstlisting}[language=fortran]
do ith_blk=0, nb_blk_th ! loop blocking in theta
 do ivpar=0, Nvpar
  call feet_computations_with_openmp(...)
!$OMP PARALLEL DO COLLAPSE(2)
   do iphi=0, iNphi
     do ith=ith_blk*th_bsize,(ith_blk+1)*th_bsize-1
       call interpolations_vectorized_kernel(...);
     end do
   end do
 end do
end do
\end{lstlisting}
\end{minipage}
}
\caption{Advection kernel external loops}
\label{protoadv_algo}
\end{minipage}\hspace*{.1cm}
\begin{minipage}[b]{9cm}
\scalebox{.55}{
\begin{minipage}{15cm}
\begin{lstlisting}[language=fortran]
#define R_BSIZE 8
subroutine interpolations_vectorized_kernel(...,spline coeff.)
 do ir_outer=0, Nr, R_BSIZE
   ! retrieve grid cell containing the foot, compute spline basis
!dir$ simd
   do ir_inner=0,R_BSIZE-1
     ir=ir_outer+ir_inner
     r_foot=...; th_foot=...; vpar_foot=...; phi_foot=...;
     ir_star       = map_on_grid(r_foot)
     ith_star      = map_on_grid(th_foot)
     ivpar_star    = map_on_grid(vpar_foot)
     iphi_star     = map_on_grid(phi_foot)
     sbasis(1:16)  = compute_spline_basis(*_star,*_foot)
   end do
   ! interpolate in combining spline basis and spline coeff.
   psum(0:R_BSIZE-1) = 0.
   do <nest_of_four_loops>
!dir$ simd
     do ir_inner=0,R_BSIZE-1
         coeff = load spline coeff. located at *_star (with unit stride)
         psum(ir_inner) = psum(ir_inner) + coeff(...) * sbasis(...)
     end do
   end do
   f1(ir_outer:ir_outer+R_BSIZE-1,ith,iphi,ivpar)=psum(0:R_BSIZE-1)
 end do
end subroutine interpolations_vectorized_kernel
\end{lstlisting}
\end{minipage}
}
\caption{Advection kernel inner loop nest}
\label{protoadv_internal}
\end{minipage}
\end{figure}


Regarding the part of the kernel presented in Figure~\ref{protoadv_algo}, the strategies applied include a blocking for the loop in the $\theta$ dimension.
It has been split in two loops (lines~1 and~6) in order to improve cache locality.
The outermost loops along $\theta$ blocks (line~1) and $\vpar$ (line~2) directions are not parallelized with \texttt{OpenMP}, instead, the loops along $\varphi$ (line~5) and inside blocks of $\theta$ (line~6) are. Thus, several data are shared between the threads that fit  into cache memory, which is not the case if parallelization would occur at outer loops.
This improves the use of cache memory (in the inner loops) which is a crucial element for the Phi. Similarly, we interlaced inside this loop nest the \textbf{feet computations} and the \textbf{interpolations} in order to reuse the data shared by these routines and to improve temporal locality.
Due to the high number of threads on Phi, getting sufficient loop counts requires the domain size along the OpenMP parallelized dimensions to be large enough.

The part of the kernel presented in Figure~\ref{protoadv_internal} corresponds to the part executed by a single thread where the parallelism comes for SIMD vectorization.
The loop along $r$ has been blocked with the iteration along $r$ blocks appearing at line~3 so as to help vectorization by having the innermost loop the same size as that of a processor vector.
This vectorization is further favored by iterating over small arrays the size of a vector such as \texttt{psum} (line~16) in the code to accumulate results inside a loop.
The upper bound of the loop along $r$ blocks (line~3) is fixed at compile time which accelerates computations on Phi a lot (not true on Sandy).
The body of the iterations inside these blocks has been split in two parts (lines~6 and~19) which enables the compiler to apply better optimizations as seen in the previous section.
The loading of spline coefficients (line~20) has also been tuned in order to read in memory with unit-stride.
Finally, the choice amongst the possible SIMD directives (\textit{i.e.} \texttt{vector} or \texttt{simd}) has been made by testing both as their impact on performance is hard to predict.

\subsection{Benchmark}
\subsubsection{Advection, simplified foot computation}
\label{bench_fake}

In order to evaluate this kernel in a way that can be compared with the previous section, we first focus on the interpolation part only (spline coefficients are computed during initialization).
To reach that goal, simplified precomputed feet are used which lead to incorrect results but induce no computing cost.
Table \ref{perf_miniapp} shows the results of a benchmark with a domain of size $N_r=128, N_\theta=128, N_\varphi=128, N_{\vpar}=64$.
80 \gf{} are obtained on Phi which is more than twice the 33 \gf{} obtained on Sandy.
Given the ratio of performance between the two architecture this shows that the optimization for the Phi are indeed well done.
Reaching this level of performance however required a huge investment in time.
The reason was that each optimization reduces execution times just a little bit (sometimes it is even hard to measure). 
But once all of these optimization are active, there is a net performance improvement.
\begin{table}[h!]
\medskip
\begin{tabular}{|ll|ll|ll|}
\hline
\multicolumn{2}{|c|}{Kernel/archi.} & \multicolumn{2}{c|}{Processing perf.} & \multicolumn{2}{c|}{Memory bandwidth}\\
\hline
Advec 4D (without foot comp.) & \textbf{Phi}  &  80 \gf{} &(7\% peak) & 2.7 \gbs{} & (2\% \textsc{stream})\\
                              & \textbf{Sandy} &  33 \gf{} & (9\% peak) &  1.1 \gbs{} & (1.6\% \textsc{stream})\\
\hline
\end{tabular}
\medskip
\caption{Performance of advection kernel (236 threads on Xeon Phi and 16 on Sandy Bridge)}
\label{perf_miniapp}
\vspace*{-0.6cm}
\end{table}
The percentage of the theoretical attainable hardware peak is however not as good as that of the kernels presented in the previous section.
This is most likely due to the integer computations and memory indirections required for this version.
In addition, all the memory accesses for the advection kernel cannot be well aligned (the foot can be located everywhere). 

It is worth noting that the present results have been obtained with a
domain size well suited to the Phi.  The performance in \gf{} is decreased for 
smaller domain sizes.  This dependency on input parameters is not as
sensitive on Sandy. The pressure and the constraints on the computing
units and the cache hierarchy seem to be higher on Phi than on Sandy
and a small imbalance can lead to severe performance issue on Phi.

Finally, additional issues come from the compiler version\footnote{We
  have used the following versions of \intel compiler for our
  experiments: 13.0.1, 14.0.1.106, 14.0.3.174, 14.0.4.211,
  15.0.0.090. Mainly, we give results in sections 4 and 5 with the
  14.0.1.106 version.}. From one version to another, the \intel
compiler does not employ the same SIMD instructions and optimizations
for a given code. We have noted that a well optimized code can suffer
a large slowdown (factor 3 has been observed) when changing the
compiler version. This sensitivity clearly advocates for writing the
kernel so as to guide more the compiler on the optimizations to apply, for example with \intel intrinsics.
This has several drawbacks but avoids
such dependency on the compiler.

\subsubsection{\gysela{} mini-app}
\label{bench_true}

By adding the actual computation of the feet of the characteristics and the spline coefficients, the code becomes comparable to the \gysela{} mini-app on Phi presented in section~\ref{framework}.
The performance of this optimized 
version are improved by an impressive factor 14 on medium test cases compared to the initial version (domain size
$128\times{}128\times{}64\times{}32$) and by a factor 8 compared to the \textit{top-down} version of section~\ref{firstminiapp}. A single advection step takes 3.2s now (45s initially, 25s after top-down enhancements). 
Compared to \textit{top-down} version, the \textit{bottom-up} approach exhibits several benefits. 
Firstly, building the kernel step-by-step enables to benchmark tiny/elementary parts of the code and to control the compiler in order to trigger the usage of deisred vectorized instructions.
Secondly, identifying block sizes for cache optimization, prefetch parameters and optimizing data loading with indirections is much easier in this context.
Thirdly, the implementation of the 3D interpolation does not start from scratch. 
The starting point is performing 1D and 2D interpolations.
Increasing algorithm complexity step by step while keeping a good level of performance was the key ingredient in this \textbf{bottom-up} approach. 

The global execution
time for the 4D advection however remains two times larger on Phi than on Sandy
(one node - two sockets), we observe that one single advection takes only 1.5s. 
This means the Phi version remains less efficient in its use of computing resources.
But, even in the previous subsection \ref{bench_fake}, we
have seen that Phi can be more than 2 times faster than Sandy on the
tuned 4D kernel alone. One can wonder why this degradation is observed ?


The problem comes from the fact that the advection time is interlaced
with (and hidden by) two other parts that require significant
computation time (the corresponding routines are not well vectorized
yet). These parts are: the computation of the feet and the spline
coefficient computations, which are both memory-bound kernels. It
would require the same kind of study that we have presented so far on
the interpolation to speedup these two supplementary parts. Another
important point is related to the temporal locality of the data
structures that are transmitted from one subroutine to the other in
the 4D advection. In that purpose, we had to ensure that the memory area that stores
the feet of the characteristics (line 3 of Algorithm~\ref{protoadv_algo})
remains in cache memory when advection is performed (line 5 to 9). But
this memory area size is difficult to tune, while the tiling sizes
also have to satisfy the very fine grain needed on Phi (this
constraint is not so hard on Sandy Bridge due to the much lower number of threads: 16 versus 236).

The optimization effort we have made to reduce the execution time on
Xeon Phi has however also been useful to lower the computational costs on
Sandy Bridge. If one considers the state of the code at the beginning and at
the end of this work, the overall execution time devoted to 4D advection scheme
(Algorithm~\ref{algo4D}) has been reduced by 30\% up to 55\% on typical
physical cases. We consider this to be an especially
interesting collateral effect of the study (Xeon Phi was targeted, not
Sandy Bridge architecture).

\subsubsection{Splitting versus no-splitting}

Let us now come back on the analysis done in 
subsection~\ref{split_scheme}. The Strang splitting scheme with four 1D
advections and one 2D advection (\textit{Split}) is
clearly memory-bound and this scheme has been used for a long time in
\gysela{}. On the other hand, the most recent 4D advection scheme
(\textit{Nosplit}) leads to a larger number of
floating-point operations but requires ideally less memory
transfers. Now that several tuning operations have been done on the 4D kernel,
the \textit{Nosplit} version has reached a quite high
optimization level on Sandy Bridge architecture.

After this work on 4D kernels, we observe on Sandy Bridge that the 
\textit{Nosplit} case is slightly faster than the \textit{Split} case.
The execution time is reduced by 1\% to 20\% depending on the
domain size. The 4D advection is now an interesting solution
because, despite its expensive computations, execution time
is better than for the splitting scheme. This gain is even larger
for large domain sizes. On the one hand, the \textit{Nosplit} case
exhibits much more compact kernels in term of the number of code
lines. This is definitely an advantage from the software 
engineering point of view. On the other hand, the \textit{Nosplit} case
also has drawbacks. First, it
requires temporary buffers to store the feet of the characteristics (extra
memory consumption). Second, tiling sizes have to be specifically 
tuned for each new machine in order to benefit from maximal cache effects.

Considering the communication costs of the \textit{NoSplit}
case in a future application using a MPI parallelization, as the 
parallel algorithm is not yet known, it is difficult to make accurate
predictions. Nevertheless, the constraints and data dependencies are 
quite similar to those of the \textit{Split} case, therefore we do not
expect that the amount of communications will be completely different.

Another point, we have also made some time measurements
of \textit{Split} and \textit{NoSplit} versions on a Sandy Bridge node
with hyper-threading activated (on a small development machine with 16
cores, apart from Helios supercomputer). We have compared the
execution times of the medium test case (domain size
$128\times{}128\times{}64\times{}32$) with 16 threads versus 32
threads. We took the best execution times among all the possible
deployments of the threads on the cores in each case.  As a result,
the hyper-threaded configuration with 32 threads shows a reduction by
9\% of the execution time (for the advection step), compared to 16
threads hyper-threading deactivated configuration. This reduction is
observed both for \textit{Split} and \textit{NoSplit} versions.



%

\section{\label{concl}Conclusion}

In this paper, we have presented the optimization of a Semi-Lagrangian Vlasov solver for the \intel{} Xeon Phi architecture.
We started this work by running several memory and network micro-benchmarks to get familiar with this new hardware and to set up the kind of performance we should expect from it.
Then, starting from the production version of \gysela{} code, several simplified versions were derived in order to explore different implementation strategies with more flexibility. 
When only the heart of the numerical method (the advection) was remaining and the expected performance could not be reached, it has been decided to start again with very simple interpolation kernels.
By increasing incrementally the code complexity, the full 4D spline interpolation kernel could be implemented.
The final obtained performance was satisfactory as the computing time was twice as fast on Xeon Phi as on a dual-socket Sandy bridge node. 
This kernel has been successfully integrated back into the \gysela{} mini-app.
But because of the other parts of the advection (feet and spline coefficients computations), the code still runs slower by a factor 2 on Xeon Phi compared to Sandy Bridge.
This is because these other parts are also difficult to vectorize/optimize/parallelize for the Xeon Phi.
Complementary work still has to be done to recover the missing factor 4 on the execution time on Xeon Phi.
To port the whole production version of \gysela{} code on such architecture remains a long term objective.

All along this optimization process, performance evaluation have been performed on both the Sandy Bridge and Phi computing units.
For the best cases of the simple kernels, we managed to reach similar fractions of the peak performance on Phi as on Sandy Bridge.
In other cases however, and especially when the code complexity builds up, we did not manage to reach that goal.

Regardless the final result, the investment in terms of code engineering to reach these performance levels is very large.
The level of parallelism to extract is much higher than on Sandy Bridge: 177 and up to 236 OpenMP threads are required on Phi versus 16 on Sandy Bridge.
Extracting this fine grain parallelism is difficult for some kernels, especially for small domain sizes.
Many other aspects have to be looked at, including vectorization, prefetching, data alignment, cache locality, cache trashing and memory access patterns while these have much less impact on Sandy Bridge.
In addition, compiler optimizations are very important to get good performance but are very difficult to predict.
Some good practice can help such as loop splitting, but no golden rule can be extracted and in practice, one often has to look at the generated assembly code to assess its quality.
This problem is further amplified by the fact that the same code can yield very different performance results depending on the compiler version.

In order to reduce this cost, we also hope for progress on the Xeon Phi hardware and software side with the next generation of processor:
\begin{itemize}
\item The hardware architecture and software stack (including MPI) should provide a more homogeneous view of the platform, reducing communication overheads.
\item An improved compiler that produces performant executables with less programming effort. This would allow to port large applications on Xeon Phi while achieving better performance.
\item Support for a richer set of directives (for example for SIMD) would enable a finer control of the vectorization process when the automatic one is not satisfactory.
\item Improvements in the hardware and OS might reduce the jitter in execution time from one run to another. 
\item The presence of a few ``Big cores'' directly on the MIC in addition to the many small cores would reduce the impact of the serial portions of the code.
\end{itemize}

A collateral benefit of this work 
is that the \gy{} miniapp performance on Sandy Bridge has increased by almost a factor two 
in some case using 4D advections. The 4D advection
approach is now faster than the classical Strang splitting scheme that is
used for a long time in
\gy{}.
This result opens perspectives for the improvement of both the numerical method precision and 
the implementation of a new parallelization strategy which would enable exascale computing 
for \gy{}.

{\footnotesize
\subsection*{Acknowledgements}
We would like to thank Thomas Guillet from \intel{} Exascale Labs for
fruitful discussions about optimization on Xeon Phi. This work has been done with
the support of High Level Support Team (IPP Garching). We have
used the computing facility of IFERC-CSC in Rokkasho and in particular
the Helios supercomputer. We would like to thank also LRZ, Cineca, Bull R\&D and 
RZG computing centers for granting us access to their Xeon Phi clusters.
Most of this project has been funded by ANR
GYPSI and G8-Exascale Nu-FuSE projects. 
}

\vspace*{-0.1cm}
\bibliographystyle{alpha}
\bibliography{gysela}

\end{document}